\documentclass[conference]{IEEEtran}
\usepackage{cite}
\usepackage{amsmath,amssymb,amsfonts}
\usepackage{algorithmic}
\usepackage{graphicx}
\usepackage{textcomp}
\usepackage{xcolor}
\usepackage{url}
\def\BibTeX{{\rm B\kern-.05em{\sc i\kern-.025em b}\kern-.08em
    T\kern-.1667em\lower.7ex\hbox{E}\kern-.125emX}}
\begin{document}

\title{POST: Email Archival, Processing and Flagging Stack for Incident Responders}

\author{\IEEEauthorblockN{Jeffrey Fairbanks}
\IEEEauthorblockA{\textit{GCFA, GDAT} \\
\textit{Lawrence Livermore National Lab}\\
Livermore, California, USA \\
fairbanks6@llnl.gov}
% \and

% % \IEEEauthorblockN{Cody Willard}
% % \IEEEauthorblockA{\textit{Computing} \\
% % \textit{Lawrence Livermore National Laboratory}\\
% % Livermore, California, USA \\
% % willard@llnl.gov}
% % }

% % \and
% \IEEEauthorblockN{Edoardo Serra}
% \IEEEauthorblockA{\textit{Computing} \\
% \textit{Boise State University}\\
% Boise, Idaho, USA \\
% edoardoserra@boisestate.edu}
}

\maketitle

%%% WHERE WE WANT TO PUBLISH THIS::
%% https://cikm2024.org/call-for-demo-papers/

\begin{abstract}

Phishing is one of the main points of compromise, with email security and awareness being estimated at \$50-100B in 2022.
 There is great need for email forensics capability to quickly search for malicious content. A novel solution POST is proposed. 
 POST is an API driven serverless email archival, processing, and flagging workflow for both large and small organizations that collects and parses all email, 
 flags emails using state of the art Natural Language Processing and Machine Learning, allows full email searching on every aspect of an email, 
 and provides a cost savings of up to 68.6\%.
\end{abstract}


\section{Introduction}

Phishing is a well-documented attack vector that has received substantial attention in literature. Research indicates that phishing attacks represent the most common form of cyber threat, yet meaningful progress has only slowly addressed this persistent challenge over time \cite{mostProminentAttacksFacedByInternetUsers}. This dynamic may suggest a degree of normalization or desensitization to the ongoing vulnerability of core IT systems and access points to such attacks within the field. Empirical studies have consistently identified phishing as the preeminent attack vector, with this trend persisting on an annual basis and email security and awareness being estimated at \$50-100B in 2022 \cite{mckinseySurveyReveals}. Furthermore, recent industry research conducted by Barracuda, a prominent provider of email protection solutions, indicates that in 2022, 73\% of surveyed organizations experienced a successful ransomware attack, with 69\% of these incidents originating through email phishing vectors \cite{barracuda_ransom}. While email gateway services play a critical role in security, their forensic capabilities are often limited, as they may not provide comprehensive searching of email bodies and attachments for malicious content at a price point that small to medium sized organizations can digest, or even at all. This points to a broader need for improved email searching functionality, as well as API-driven solutions enabling full email parsing, searching, and flagging capabilities at low cost. Such enhancements could bolster endpoint protection, automation, and machine learning-based security measures. The present disclosure, POST, relates to a system and method for email Parsing, Searching, Flagging and Archival services. POST is completely lightweight and serverless in its implementation, allowing for a fully templateized architecture that can be quickly implemented ranging in small organizations to large organizations. Through this cloud native serverless approach, the cost of running full functionality within POST is substantially lower than others in this space, while still allowing room for state-of-the-art NLP and Machine learning tasks to be completed on every email, both through dynamic and static analysis.

% talk about Machine learning LLM Approaches being too general 
In response to the limitations of existing email security solutions, the present approach adopts a proactive methodology leveraging state-of-the-art large language models, machine learning, and other natural language processing techniques to conduct a granular, multi-faceted inspection of each email. 
% Facilitated by a serverless architecture, this framework enables the application of finely tuned machine learning and language model configurations tailored to the analysis of discrete email components.
In contrast to more generalized machine learning approaches common in this domain, the present solution undertakes a detailed, targeted examination of email contents. This specialized, component-level analysis represents a departure from conventional strategies, promising enhanced detection and mitigation of evolving email-borne threats.

\section{System Overview}

\begin{figure*}[t!]
    \centering
    \includegraphics[width=\textwidth]{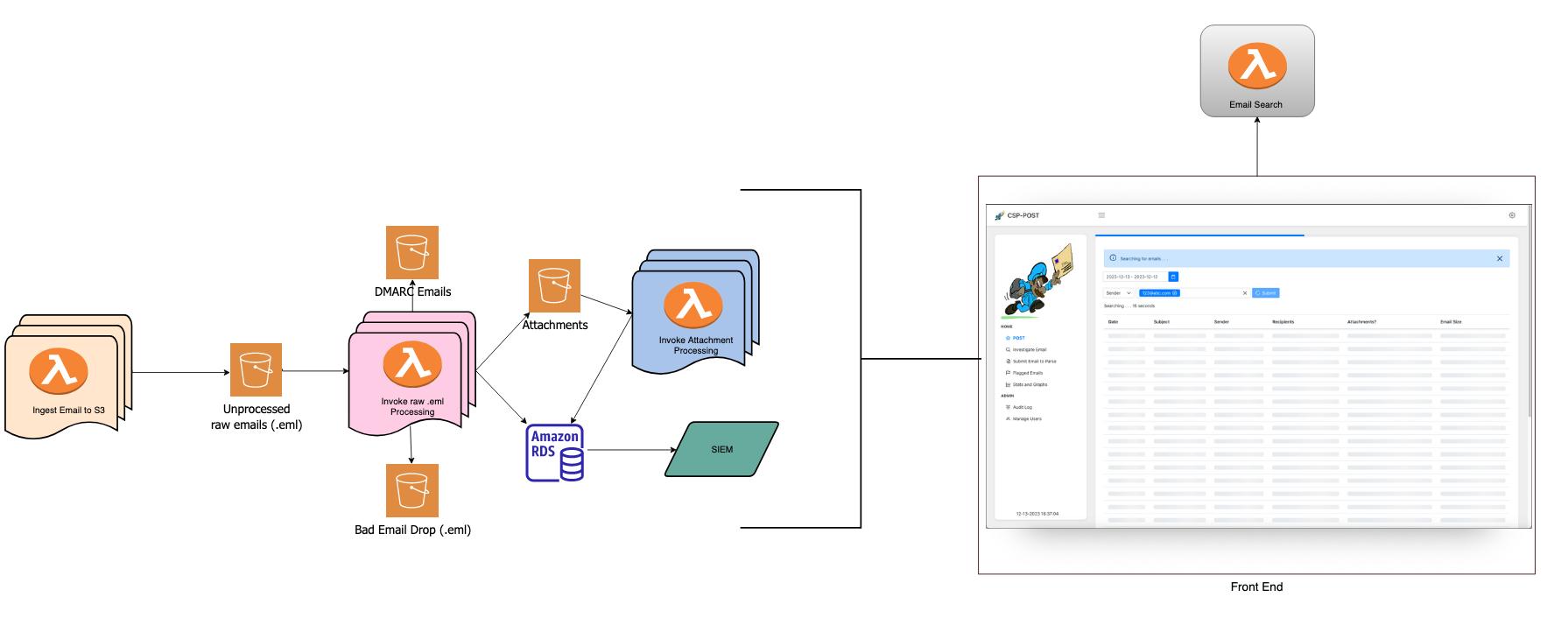}
    \caption{POST Overview}
    \label{fig:figure1}
\end{figure*}

The present workflow adopts an event-driven architecture, facilitating seamless operation without the need for additional worker processes or background daemons. The system encompasses two core functionalities. First, it ingests and parses all incoming and outgoing emails, extracting easily searchable metadata. Second, it provides robust forensic capabilities through comprehensive email inspection and flagging. This system and method may be deployed on a cloud-based platform, such as AWS CloudFormation, in a repeatable, testable, and auditable manner. In one aspect, the present disclosure describes a system for email parsing and archival services. The system comprises an ingestion module that ingests emails from a source system into a database, and a parsing module that extracts data from the ingested emails, including metadata, links, email bodies, and attachments. 
In another aspect, the disclosure outlines a method for deploying a processing module capable of statically and dynamically extracting flags from email data, flagging emails based on malicious content, and submitting them to a bulk file analysis framework.

The present solution is designed to scale based on incoming and outgoing email volume, enabling both small and large organizations to leverage its full functionality while maintaining cost-effectiveness and operational simplicity. Preliminary analysis indicates that this approach can deliver cost savings of up to 68.6\% compared to alternative solutions \cite{proofpoint}.

% A critical part of the architecture is the provision of API endpoints for every aspect of the workflow, enabling users to integrate the system with existing frontends, incorporate it into custom machine learning and analysis pipelines, or leverage it as a core technology for other applications. This modular design allows organizations to selectively enable or disable specific components without disrupting the overall functionality.

The use of a templatized, serverless approach, coupled with the API-driven architecture, facilitates one-click deployment of the entire workflow within an organization. This streamlines setup and configuration, allowing the Incident Response and Digital Forensics teams to focus on core security tasks, such as threat hunting, incident response, and phishing awareness, rather than time-consuming infrastructure management. Additionally, POST provides comprehensive email searching capabilities, including within email bodies and attachments - a feature often associated with high computational and temporal costs. Users can leverage this functionality to perform targeted, after-the-fact searches for keywords and other indicators of compromise. Finally, POST incorporates robust email flagging capabilities powered by state-of-the-art large language models and machine learning techniques. These models inspect incoming and outgoing emails for indicators of phishing, spam, and insider threats, analyzing artifacts such as HTML, links, and QR codes to proactively detect and mitigate evolving email-based attacks.

Figure 1 depicts the main architectural overview of the proposed workflow. The process commences with the Ingestion module, which collects all incoming and outgoing email communications. Once an email is ingested into the storage system, the subsequent "Raw .EML Processing" workflow is initiated. This workflow manages the parsing of email data, conducts static analysis on the .eml files, and performs both traditional machine learning and large language model-based natural language processing tasks. If the ingested email contains attachments, those files are stored in a separate storage bucket, triggering the "Attachment Processing" workflow. This secondary workflow undertakes static and dynamic analysis directly on the attachment data. Furthermore, it provides the capability to submit each attachment to a bulk analysis framework, such as a dynamic sandbox environment, for comprehensive inspection. All the extracted data, including metadata, email bodies, and attachment information, is stored in a PostgreSQL database, enabling efficient searching and querying through API endpoints or a frontend interface. Additionally, any flags or indicators of compromise identified through the machine learning and language model-based analyses are fed into the organization's Security Information and Event Management (SIEM) system, supporting the efforts of the Digital Forensics and Incident Response teams.

\subsection{Ingest}

The proposed system, referred to as POST, offers flexible email ingestion capabilities to accommodate various architectural considerations within the target environment. The system provides built-in functionality to ingest emails directly from the organization's SMTP server. 
% In this implementation, AWS Lambda functions are utilized to periodically (e.g., on a 1-minute cron-like schedule) poll the SMTP server and transfer all new emails to the designated S3 storage bucket. 
% These Lambda functions are designed to run for up to 15 minutes, ensuring the complete transfer of all available email data.

% % \begin{figure}[htbp]
% %     \centering
% %     \includegraphics[width=0.3\textwidth]{figures/POST-ingest.png}
% %     \caption{Utilizing AWS Lambda}
% %     \label{fig:figure3}
% % \end{figure}

% Once the emails have been ingested and stored in the S3 bucket, the POST workflow initiates the subsequent processing tasks, which may include parsing, analysis, and other operations as defined by the system configuration.

\subsection{Raw .eml Processing}

The "Raw .EML Processing" segment of POST ensures the completion of essential tasks associated with .eml files. This includes parsing email content to facilitate efficient and targeted searching, extracting embedded files and attachments for downstream analysis, and executing various machine learning-based evaluation of the raw .eml data. Through the execution of these processing tasks, each ingested email is indexed and stored in a manner that enables efficient searching and retrieval at a later stage. Additionally, the analyses conducted on the raw .eml data facilitate the identification and flagging of any malicious or spam-related indicators present within the email contents. Through this flagging, the email can be pulled from the users inbox, or blocked upon arrival.

% By systematically addressing these core processing steps, the system establishes a foundation for comprehensive email data extraction, storage, and analysis to support various security and investigative use cases.

% \begin{figure}[htbp]
%     \centering
%     \includegraphics[width=0.3\textwidth]{figures/POST_RawEmlProcessing.png}
%     \caption{POST Raw Eml Processing Tasks}
%     \label{fig:figure4}
% \end{figure}

\subsubsection{parsing}

The parsing function extracts all metadata from the ingested emails and persists this information either in an AWS RDS PostgreSQL database or various S3 data storage buckets. This parsing operation comprehensively extracts all relevant components of the email content. To handle the individual emails arriving in the S3 bucket, a dedicated AWS Lambda function is spawned for each item. Multiple such parsing Lambdas can execute concurrently, each processing a distinct .eml file. The Lambda function is responsible for retrieving all pertinent data from the email and storing it in a manner that enables subsequent searching and indexing. This extracted data, residing in the S3 buckets and RDS database, is then accessible through search-oriented Lambda functions and other API endpoints for further analysis and investigation. By adopting this approach, the system establishes a scalable, distributed email parsing and storage architecture, facilitating efficient retrieval and querying of the ingested email data to support security and forensic use cases.

\subsubsection{static analysis}

The "Raw .EML Processing" workflow also incorporates static analysis functionalities that evaluate the .eml content directly, without consideration of any attached files. The inspection and analysis of email attachments is delegated to the separate "Attachment Processing" workflow. The static analysis component subjects the emails to various scanning techniques, including the utilization of the YARA rule-based pattern matching system. YARA is effective in detecting known signatures and facilitates the sharing of prominent indicators of compromise (IOC) across the security community \cite{cisa_yara} \cite{lockett2021assessing}. Additionally, this workflow analyzes the URLs, IP addresses, sender information, and recipient details within the emails, cross-referencing them against the organization's existing block lists and other relevant threat intelligence sources. By executing these static analysis tasks on the raw .eml files, the system is able to identify and flag indicators of malicious activity present within the email content itself, prior to any attachment-based inspection. This approach helps establish a comprehensive security posture by addressing threats that may manifest in the email body, headers, and metadata. This static analysis functionality helps identify and mitigate lower-level attack vectors, serving as an initial filtering mechanism to address more readily detectable threats.

\subsubsection{Machine learning and NLP tasks}

The core of the flagging, blocking, and incident response capabilities within the POST system lies in the application of machine learning and natural language processing techniques, including the utilization of large language models (LLMs). These advanced analytical components provide the system with the ability to classify emails into categories of benign, spam, or malicious content. This classification allows for the appropriate remediation actions, such as removing malicious emails from the organization, facilitating cleanup efforts, and bolstering protection against evolving spear-phishing attacks. The robust integration of these machine learning and NLP-powered functionalities represents a fundamental aspect of the POST system's comprehensive approach to email security and threat mitigation. By leveraging state-of-the-art techniques in automated email analysis and classification, the system delivers enhanced capabilities to support the organization's overall incident response and security operations.

One of the key approaches employed within the POST system involves the use of Random Forest classifiers to analyze the email body content and categorize messages as malicious, spam, or benign. This technique builds upon the extensive body of research and prior work in the field of automated email classification \cite{8252051}. However, it is important to note that this Random Forest based classification represents just one component within the broader suite of processing and flagging operations conducted on the .eml files. The system leverages a multifaceted approach, incorporating this well-established technique as part of a larger, comprehensive analysis framework.

Additionally, the system leverages large language models (LLMs) to analyze the email body content, with the objective of identifying and mitigating social engineering schemes and spear-phishing attacks. In high-risk environments, sophisticated threat actors frequently leverage spear-phishing techniques \cite{cisaSophisticatedSpearphishing}. This has prompted the preliminary application of LLM-based approaches to detect social engineering and spear-phishing within the present system, POST. The LLM models are trained to analyze sender-recipient relationships and historical email thread interactions, utilizing retrieval augmentation techniques to build context and identify anomalous patterns indicative of such attacks. The successful demonstration and validation of this LLM-powered approach through the POST system will be the subject of future research publications, further contributing to the evolving body of knowledge in this domain.

\subsection{Attachment Processing}

The "Attachment Processing" segment of the POST workflow is dedicated to handling and extracting indicators of compromise (IoCs) from the email attachment data. This specialized functionality ensures the comprehensive analysis of any files or documents associated with the ingested emails. By addressing the attachment-based threats as a distinct component of the overall workflow, the system is able to apply targeted techniques and methodologies tailored to the inspection and identification of malicious artifacts within these supplementary email elements. 

% COMBINE THE ASSEMBLYLINE INTO THIS PICTURE!!!!
% \begin{figure}[htbp]
%     \centering
%     \includegraphics[width=0.3\textwidth]{figures/POST_AttachmentProcessing.png}
%     \caption{POST Attachment Processing}
%     \label{fig:figure5}
% \end{figure}

There are three key functionalities in this section: QR code analysis, detection and analysis of obfuscated JavaScript, and submission of files to a bulk malware sandbox environment.
First, the system examines attachments to determine the presence of QR codes. As QR code-based attacks have become increasingly prevalent in recent years \cite{ftcScammersHide}, the solution analyzes any discovered QR codes to assess the maliciousness of their associated redirects. This approach ensures the system can effectively mitigate threats that leverage QR codes to obfuscate malicious links, which may otherwise evade detection through traditional email body analysis. Second, the system utilizes Random Forest classifiers to detect the presence of HTML and JavaScript content within attachments. These advanced analytical techniques are employed to identify obfuscated scripts, which are often associated with credential harvesting attacks. By exposing and inspecting such obfuscated code, the system can effectively flag and remediate these attachment-based attack vectors. Finally, all email attachments are submitted to the Assemblyline scalable file triage and malware analysis system. Assemblyline integrates a range of community-sourced tools to dynamically analyze the attachments within a dedicated malware sandbox environment. This comprehensive attachment-focused inspection enables the detection of even the most sophisticated malicious payloads \cite{cybercentrecanadaAssemblyline}

% By addressing these key attachment-based attack vectors, the system enhances the overall security posture and threat mitigation capabilities within the email processing workflow.

\section{Demonstration}

This section introduces the key functionalities of POST.
A fundamental architectural design principle of the POST system is the implementation of every action and function as a dedicated API endpoint. This API-driven modular approach enables users to leverage the full suite of capabilities through command-line interfaces, software engineering workflows, and other programmatic integration, in addition to the web-based frontend.

% \begin{figure}[htbp]
%     \centering
%     \includegraphics[width=0.8\textwidth]{figures/POST_frontend_Search.png}
%     \caption{Demonstraing a search in POST}
%     \label{fig:figure7}
% \end{figure}

The search functionality provides users with a streamlined interface to query the system. Searches can be conducted across a wide range of email attributes, including the message body and attachment data, returning relevant results based on the specified criteria. This comprehensive search capability empowers analysts to efficiently locate and investigate emails of interest. Upon selecting a specific search result, users can drill down into the detailed information associated with that email. This includes the system's assessment of the email's threat rating, whether it was successfully delivered to the recipient or blocked, and the rationale behind any flagging or classification decisions made by the various processing components. This level of transparency and audit-ability is crucial for supporting digital forensics and incident response workflows.

% \begin{figure}[htbp]
%     \centering
%     \includegraphics[width=0.8\textwidth]{figures/POST_frontend_inspect.png}
%     \caption{Investigating an email in POST}
%     \label{fig:figure8}
% \end{figure}

A key advantage of the POST system is its ability to provide enterprise-grade email security and forensic functionality at a comparatively lower cost, making it accessible even to smaller and medium-sized organizations. This cost-effectiveness, coupled with the comprehensive API-driven architecture, allows these entities to leverage state-of-the-art email analysis and threat detection capabilities without the burden of managing complex infrastructure or incurring exorbitant upfront investments. As Cybersecurity requires ongoing investment and maintenance, a primary concern when evaluating tools in this domain is often the prohibitive cost, with he financial burden associated with many established email security solutions becoming a significant barrier to adoption \cite{designrushCostCybersecurity}. In contrast, preliminary analyses indicate that the cost of deploying the present POST system at scale is up to 68.6 times lower than the expense of utilizing a popular email gateway solution \cite{proofpoint}. This substantial cost differential is a critical consideration for organizations seeking to enhance their email security posture without incurring unsustainable financial overhead.

For instance, to support a user base of 10,000 generating 18 million emails per month, another leading email gateway solution would cost the organization an estimated \$823,200 annually according to the vendor's published price list \cite{proofpoint}. By comparison, deploying the POST system at the same scale would cost a mere \$12,000 per year, providing full functionality. By offering a more cost-effective solution without compromising on capabilities, the POST system represents a compelling alternative that may be more accessible to a wider range of stakeholders, including resource-constrained organizations. This favorable cost-benefit ratio can enable smaller and medium-sized entities to bolster their cybersecurity defenses against evolving email-borne threats.

\section{Related work And Novelty}

The Laika system, developed by Lockheed Martin, is an object scanner and intrusion detection tool with powerful implementation in the cybersecurity domain \cite{osti_1762125}. Laika's ability to rapidly analyze files has enabled its scalable deployment, particularly in high-volume environments. However, the system's architecture, which has been in production since 2012, exhibits a degree of complexity, requiring multiple worker processes and daemons to operate. While Laika adopts a modular approach akin to the proposed POST, its primary focus remains on static analysis through the utilization of YARA rules. In contrast, POST aims to advance beyond this limited scope by incorporating a broader range of static and dynamic analysis techniques.

% https://github.com/sublime-security/sublime-platform
% also talk about this tool which is similar 

Recent research conducted by J.P. Morgan Chase has highlighted the effectiveness of large language models (LLMs) in email spam detection, particularly in few-shot scenarios where limited training data is available. The Spam-T5 model proposed in this work outperformed traditional baseline methods, achieving an average F1 score of 0.7498 \cite{labonne2023spamt5}. However, the authors note that the deployment of LLMs in real-world applications will require efforts to reduce their computational requirements. POST mitigates these challenges through the use of retrieval-augmented generation (RAG) and continuous training on current organizational data. Through the use of RAG, LLMs are seeing substantial increases in accuracy and time-sensitive output, without abundant computational requirements \cite{li2024enhancing}.
inquiries.

% Furthermore, the POST system draws inspiration from the PhishBERT model, which leverages a large-scale pretrained deep transformer network for phishing URL detection [5]. While PhishBERT demonstrated promising results, its sole focus on URL-based threats neglects the broader range of malicious content that may be present within email communications. In contrast, the POST system implements large-scale pretrained models capable of detecting a wider variety of email-borne threats beyond just URLs.

In the commercial email security market, vendors such as Proofpoint, Barracuda, and Cloudflare offer various solutions. However, these products often come at a significantly higher cost compared to the functionality provided by the POST system. Additionally, the POST system boasts a completely serverless, lightweight implementation, allowing for seamless integration within any organization's environment through the use of a single deployment template. By providing digital forensics and incident response teams with access to a wide range of machine learning, natural language processing, and LLM-powered email classification and search capabilities, the POST system aims to deliver a cost-effective and comprehensive solution to address the evolving email security landscape \cite{proofpoint} \cite{barracuda} \cite{cloudflareSimplifyingSaaS}.

\section*{Aknowledgements}

This work was performed under the auspices of the U.S. Department of Energy by Lawrence Livermore National Laboratory under Contract DE-AC52-07NA27344.
% LLNL has filed for patent protection on this invention. 
For further information or questions please reach out to fairbanks6@llnl.gov. \cite{llnlCSPPOSTInnovation}

\bibliographystyle{acm}
\bibliography{refs}

\begin{thebibliography}{10}

\bibitem{barracuda}
barracuda.com.
\newblock \url{https://www.barracuda.com/?utm_source=google&utm_medium=search_cpc&utm_campaign=387189501&utm_adgroup=24415181661&utm_term=barracuda&utm_position=&utm_matchtype=e&utm_device=c&utm_content=355185958226&_bt=355185958226&_bk=barracuda&_bm=e&_bn=g&_bg=24415181661&clicktype=&gad_source=1&gclid=CjwKCAjwrIixBhBbEiwACEqDJR1ewY0IZ9dilw-NoWlnCpWYOAGDM_4dZxhg1T1MAmAGSfTmKBqYzRoCzSkQAvD_BwE}.
\newblock [Accessed 19-04-2024].

\bibitem{barracuda_ransom}
barracuda.com.
\newblock \url{https://www.barracuda.com/reports/ransomware-insights-report-2023}.
\newblock [Accessed 19-04-2024].

\bibitem{mostProminentAttacksFacedByInternetUsers}
{\sc Basit, A., Zafar, M., Liu, X., Javed, A.~R., Jalil, Z., and Kifayat, K.}
\newblock A comprehensive survey of ai-enabled phishing attacks detection techniques.
\newblock {\em Telecommunication Systems 76}, 1 (2021), 139--154.

\bibitem{cisaSophisticatedSpearphishing}
{S}ophisticated {S}pearphishing {C}ampaign {T}argets {G}overnment {O}rganizations, {I}{G}{O}s, and {N}{G}{O}s | {C}{I}{S}{A} --- cisa.gov.
\newblock \url{https://www.cisa.gov/news-events/cybersecurity-advisories/aa21-148a}.
\newblock [Accessed 19-04-2024].

\bibitem{cisa_yara}
cisa.gov.
\newblock \url{https://www.cisa.gov/sites/default/files/FactSheets/NCCIC%20ICS_FactSheet_YARA_S508C.pdf}.
\newblock [Accessed 19-04-2024].

\bibitem{cloudflareSimplifyingSaaS}
{S}implifying {S}aa{S} --- cloudflare.com.
\newblock \url{https://www.cloudflare.com/lp/simplify-saas-email/?utm_source=google&utm_medium=cpc&utm_campaign=ao-fy-acq-namer_en_na-umbrella-ge-eb-prospecting-sch_g_generic_beta&utm_content=Beta_ContentOffers_ZeroTrust_EmailSecurity&utm_term=business+email+security&campaignid=71700000111199035&adgroupid=58700008431985221&creativeid=661219386415&&_bt=661219386415&_bk=business%20email%20security&_bm=b&_bn=g&_bg=153570103641&_placement=&_target=&_loc=9032016&_dv=c&awsearchcpc=1&gad_source=1&gclid=CjwKCAjwrIixBhBbEiwACEqDJeVSbLpm1brg3ISl-srJN_Kq4Vgezqr_eUXo77lwm0pfQkutJ6xpKxoCVkkQAvD_BwE&gclsrc=aw.ds}.
\newblock [Accessed 19-04-2024].

\bibitem{cybercentrecanadaAssemblyline}
{A}ssemblyline 4 --- cybercentrecanada.github.io.
\newblock \url{https://cybercentrecanada.github.io/assemblyline4_docs/}.
\newblock [Accessed 19-04-2024].

\bibitem{designrushCostCybersecurity}
{T}he {C}ost of {C}ybersecurity and {H}ow to {B}udget for {I}t --- designrush.com.
\newblock \url{https://www.designrush.com/agency/cybersecurity/trends/cost-of-cybersecurity}.
\newblock [Accessed 19-04-2024].

\bibitem{ftcScammersHide}
{S}cammers hide harmful links in {Q}{R} codes to steal your information --- consumer.ftc.gov.
\newblock \url{https://consumer.ftc.gov/consumer-alerts/2023/12/scammers-hide-harmful-links-qr-codes-steal-your-information}.
\newblock [Accessed 19-04-2024].

\bibitem{labonne2023spamt5}
{\sc Labonne, M., and Moran, S.}
\newblock Spam-t5: Benchmarking large language models for few-shot email spam detection, 2023.

\bibitem{li2024enhancing}
{\sc Li, J., Yuan, Y., and Zhang, Z.}
\newblock Enhancing llm factual accuracy with rag to counter hallucinations: A case study on domain-specific queries in private knowledge-bases.
\newblock {\em arXiv preprint arXiv:2403.10446\/} (2024).

\bibitem{llnlCSPPOSTInnovation}
{C}{S}{P}-{P}{O}{S}{T} | {I}nnovation and {P}artnerships {O}ffice --- ipo.llnl.gov.
\newblock \url{https://ipo.llnl.gov/technologies/it-and-communications/csp-post}.
\newblock [Accessed 19-04-2024].

\bibitem{lockett2021assessing}
{\sc Lockett, A.}
\newblock Assessing the effectiveness of yara rules for signature-based malware detection and classification, 2021.

\bibitem{mckinseySurveyReveals}
{N}ew survey reveals \$2 trillion market opportunity for cybersecurity technology and service providers --- mckinsey.com.
\newblock \url{https://www.mckinsey.com/capabilities/risk-and-resilience/our-insights/cybersecurity/new-survey-reveals-2-trillion-dollar-market-opportunity-for-cybersecurity-technology-and-service-providers}.
\newblock [Accessed 19-04-2024].

\bibitem{proofpoint}
proofpoint.com.
\newblock \url{https://www.proofpoint.com/sites/default/files/data-sheets/pfpt-us-ds-essentials-pricelist-msrp-us.pdf}.
\newblock [Accessed 19-04-2024].

\bibitem{8252051}
{\sc Subasi, A., Molah, E., Almkallawi, F., and Chaudhery, T.~J.}
\newblock Intelligent phishing website detection using random forest classifier.
\newblock In {\em 2017 International Conference on Electrical and Computing Technologies and Applications (ICECTA)\/} (2017), pp.~1--5.

\bibitem{osti_1762125}
{\sc Walkup, G., Walkup, E., Smutz, C., Nebergall, C., Lee, W., Gallagher, J., Coia, K., Greene, D., Healy, B., Flickinger, J., Gurule, K., Martin, L., and USDOE}.
\newblock Laika boss (binary object scanning system), version 2.00+2821.g2403134, 11 2020.

\end{thebibliography}

\vspace{12pt}
\end{document}